\documentclass[referee]{raa}
\usepackage{graphicx,times}
\usepackage{natbib}
\usepackage{amssymb,amsmath}
\bibpunct{(}{)}{;}{a}{}{,}

\usepackage[a4paper=true,dvipdfm=true,pagebackref=true]{hyperref}
\hypersetup{pdftitle = The title of my PDF, pdfauthor = My name, pdfsubject= The subject, pdfkeywords = keyword1 keyword2 keyword3}
\hypersetup{colorlinks = true, linkcolor = green, anchorcolor = red, citecolor = blue, filecolor = red, pagecolor = red, urlcolor = red}

\begin{document}

   \title{Combined Limit on the Photon Mass with Nine Localized Fast Radio Bursts}

 \volnopage{Vol.0 (20xx) No.0, 000--000}      %%preserved for Editor. DOn't remove!
   \setcounter{page}{1}

   \author{Jun-Jie Wei\inst{1,2} \and Xue-Feng Wu\inst{1,2}
   }
%% Here is an example of three authors come from different institutes.
%% For single author or all the authors from an institute, use "\inst{}" only

   \institute{Purple Mountain Observatory, Chinese Academy of Sciences, Nanjing 210023, China; {\it jjwei@pmo.ac.cn; xfwu@pmo.ac.cn}\\
	\and School of Astronomy and Space Sciences, University of Science and Technology of China, Hefei 230026, China\\
\vs \no
   {\small Received~~20xx month day; accepted~~20xx~~month day}
}

\abstract{A nonzero-mass hypothesis for the photon can produces a frequency-dependent dispersion
of light, which results in arrival-time differences of photons with different frequencies
originating from a given transient source. Extragalactic fast radio bursts (FRBs), with their
low frequency emissions, short time durations, and long propagation distances, are excellent
astrophysical probes to constrain the rest mass of the photon $m_{\gamma}$. However, the
derivation of a limit on $m_{\gamma}$ is complicated by the similar frequency dependences
of dispersion expected from the plasma and nonzero photon mass effects. If a handful
measurements of redshift for FRBs are available, the different redshift dependences of the plasma
and photon mass contributions to the dispersion measure (DM) might be able to break dispersion
degeneracy in testing the photon mass. For now, nine FRBs with redshift measurements have been
reported, which can turn this idea into reality. Taking into account the DM contributions from
both the plasma and a possible photon mass, we use the data on the nine FRBs to derive a combined
limit of $m_{\gamma}\leq7.1\times10^{-51}\;{\rm kg}$, or equivalently
$m_{\gamma}\leq4.0\times10^{-15}\; {\rm eV}/c^{2}$ at 68\% confidence level, which is essentially
as good as or represents a factor of 7 improvement over previous limits obtained by the single FRBs.
Additionally, a reasonable estimation for the DM contribution from the host galaxy, $\rm DM_{host}$,
can be simultaneously achieved in our analysis. The rapid progress in localizing FRBs will further
tighten the constraints on both $m_{\gamma}$ and $\rm DM_{host}$.
\keywords{radio continuum: general --- intergalactic medium --- astroparticle physics}
}

   \authorrunning{Wei \& Wu}            %author_head in even pages
   \titlerunning{Combined Photon Mass Limit from Nine Localized FRBs}  % title_head in odd pages
   \maketitle

%________________________________________________ sections below
%
\section{Introduction}           %% first-level sections will be auto-capitalized
\label{sect:intro}

The postulate of the ``constancy of light speed'' is a major pillar of Maxwell's
electromagnetism and Einstein's theory of special relativity. This postulate implies that
the photon, the fundamental quanta of light, should be massless. Nevertheless,
several theories have challenged the concept of the massless photon, starting from
the famous de Broglie-Proca theory \citep{deb22,pro36} and nowadays proposing models with
massive photons for dark energy (e.g., \citealt{2016PhRvD..93h3012K}). Within that context,
determining the rest mass of the photon has historically been an effective way to test the
validity of this postulate. According to the uncertainty principle, however, it is impossible
to perform any experiment that would fully prove that the photon rest mass is exactly zero.
Adopting the age of the universe of about $10^{10}$ years, the ultimate upper bound on
the photon rest mass would be $m_{\gamma}\leq\hbar/\Delta tc^2\approx10^{-69}~\rm{kg}$ (e.g.,
\citealt{1971RvMP...43..277G,2005RPPh...68...77T}). The best experimental strategy is therefore
to place ever stricter upper limits on $m_{\gamma}$ and push the experimental limits more
closely towards the ultimate upper limit.
%verification of the zero-mass hypothesis for the photon as far as possible.

From a theoretical perspective, a nonzero photon mass can be incorporated into electromagnetism
straightforwardly via the Proca equations, which are the relativistic generalization of Maxwell's
equations. Using the Proca equations, it is possible to consider some observable effects
associated with massive photons. One can then constrain the photon rest mass by searching for
such effects (see \citealt{1971RvMP...43..277G,1973PhRvD...8.2349L,2005RPPh...68...77T,
2006AcPPB..37..565O,2010RvMP...82..939G,2011EPJD...61..531S} for reviews). Based on such effects,
various experimental methods have been developed to set upper limits on the photon rest mass, such as
Coulomb's inverse square law \citep{1971PhRvL..26..721W}, Amp$\rm \grave{e}$re's law
\citep{1992PhRvL..68.3383C}, torsion balance \citep{1998PhRvL..80.1826L,2003PhRvL..90h1801L,
2003PhRvL..91n9102L}, gravitational deflection of electromagnetic waves \citep{1973PhRvD...8.2349L,
2004PhRvD..69j7501A}, Jupiter's magnetic field \citep{1975PhRvL..35.1402D}, magnetohydrodynamic
phenomena of the solar wind \citep{1997PPCF...39...73R,2007PPCF...49..429R,2016APh....82...49R},
mechanical stability of the magnetized gas in galaxies \citep{1976UsFiN.119..551C},
stability of plasma in Coma cluster \citep{2003PhRvL..91n9101G},
black hole bombs \citep{2012PhRvL.109m1102P}, pulsar spindown \citep{2017ApJ...842...23Y},
the frequency dependence in the velocity of light \citep{1964Natur.202..377L,1969Natur.222..157W,
1999PhRvL..82.4964S,2016ApJ...822L..15W,2016PhLB..757..548B,2017PhLB..768..326B,
2016JHEAp..11...20Z,2017PhRvD..95l3010S,2017RAA....17...13W,2018JCAP...07..045W,2019ApJ...882L..13X},
and so on. Among these methods, the most immediate and robust one is to measure
the frequency-dependent dispersion of light.

If the photon rest mass is nonzero, the dispersion relation is given by
\begin{equation}\label{eq1}
E=\sqrt{p^2c^2+m_\gamma^2c^4}\;.
\end{equation}
The massive photon group velocity then takes the form:
\begin{equation}\label{eq2}
\upsilon=\frac{\partial{E}}{\partial{p}}=c\sqrt{1-\frac{m_\gamma^2c^4}{E^2}}\approx c\left(1-\frac{1}{2}\frac{m_\gamma^2c^4}{h^2\nu^2}\right)\;,
\end{equation}
where $\nu$ is the frequency and the last approximation holds when
$m_\gamma\ll h\nu/c^{2}\simeq7\times10^{-42}\left(\frac{\nu}{\rm GHz}\right)\;{\rm kg}$.
Equation~(\ref{eq2}) means that low frequency photons propagate in a vacuum slower than
high frequency ones. Two photons with different frequencies, if emitted simultaneously
from the same source, would reach us at different times. Moreover, it is obvious that
observations of shorter time structures in lower energy bands from astronomical sources
at cosmological distances are particularly sensitive to the photon mass. Thanks to
their short time durations, long propagation distances, and low frequency emissions,
extragalactic fast radio bursts (FRBs) provide the most promising celestial laboratory
so far for constraining the photon mass through the dispersion method
\citep{2016ApJ...822L..15W,2016PhLB..757..548B,2017PhLB..768..326B,2017PhRvD..95l3010S,2019ApJ...882L..13X}.
For instance, taking the controversial redshift measurement of FRB 150418 ($z=0.492$),
\footnote{The redshift identification of FRB 150418 has been challenged with
an active galactic nucleus variability \citep{2016ApJ...824L...9V,2016ApJ...821L..22W},
and is now generally not accepted \citep{2017Natur.541...58C}.}
\cite{2016ApJ...822L..15W} set a tight upper limit on the photon mass as
$m_{\gamma}\leq5.2\times10^{-50}~\rm{kg}$, which is three orders of magnitude better than
the previous best constraint from other astronomical sources with the same method
(see also \citealt{2016PhLB..757..548B}). Later, \cite{2017PhLB..768..326B} used the
reliable redshift measurement of FRB 121102 to constrain the photon mass to be
$m_{\gamma}\leq3.9\times10^{-50}~\rm{kg}$. Using the time-frequency structure of
subpulses in FRB 121102, \cite{2019ApJ...882L..13X} obtained a stricter limit on
the photon mass of $m_{\gamma}\leq5.1\times10^{-51}~\rm{kg}$.

From observations, the pulse arrival times of FRBs strictly follow the $1/\nu^{2}$ law,
which is in good agreement with the propagation of radio signals through a cold plasma.
However, a similar frequency-dependent dispersion $\propto m_\gamma^2/\nu^2$ (see Equation~\ref{eq2})
could also be resulted from a nonzero photon mass. The similarity between the frequency-dependent
dispersions induced by the plasma and nonzero photon mass effects complicates the derivation of
a constraint on $m_{\gamma}$. Fortunately, since the two dispersions have different dependences
on the redshift $z$, it has been suggested that they could in principle be distinguished
when a handful redshift measurements of FRBs become available, and making the sensitivity to
$m_{\gamma}$ to be improved \citep{2016PhLB..757..548B,2017PhLB..768..326B,2017AdSpR..59..736B}.
To date, nine FRBs, including FRB 121102 \citep{2016Natur.531..202S,2017Natur.541...58C,2017ApJ...834L...7T},
FRB 180916.J0158+65 \citep{2020Natur.577..190M}, FRB 180924 \citep{2019Sci...365..565B},
FRB 181112 \citep{2019Sci...366..231P}, FRB 190523 \citep{2019Natur.572..352R},
FRB 190102, FRB 190608, FRB 190611, and FRB 190711 \citep{2020Natur.581..391M},
have already be localized. With the measurements of nine FRB redshifts in hand, it is
interesting to investigate what level of photon mass limits can be improved by taking
advantage of the different redshift dependences of the dispersions from the plasma and
photon mass. In this work, we develop a statistical approach to obtaining a combined limit
on $m_{\gamma}$ using the nine FRBs with redshift measurements. In addition, such an approach
can also simultaneously lead to an estimate of the mean value of the dispersion measure (DM)
contributed from the local host galaxy.

We should note that a recent work by \cite{2017PhRvD..95l3010S} extended previous studies to
those FRBs without redshift measurement. They used an uninformative prior for the redshift
to derive a combined constraint on $m_{\gamma}$ from a catalog of 21 FRBs. Comparison of
our results will be interesting, because the exact value of the redshift (or distance)
plays an important role in constraining the photon mass. It is therefore useful to
crosscheck, especially as using the real redshift measurements of FRBs, as we do in this work,
avoids potential bias from the prior of redshift.

This paper is arranged as follows. In \S~\ref{sec:method}, we introduce the analysis method
used to disentangle the dispersions from the plasma and photon mass and to constrain the
photon mass. A combined limit on the photon mass from nine localized FRBs will be presented
in \S~\ref{sec:result}, followed by our conclusions in \S~\ref{sec:summary}.
Throughout this paper, we adopt the flat $\Lambda$CDM model with the cosmological parameters
recently derived from the latest $Planck$ observations \citep{2018arXiv180706209P}:
the Hubble constant $H_0=67.36$ km $\rm s^{-1}$ $\rm Mpc^{-1}$, the matter energy density
$\Omega_{\rm m}=0.315$, the vacuum energy density $\Omega_{\Lambda}=0.685$, and the baryonic
matter energy density $\Omega_{b}=0.0493$.

\section{Analysis method}
\label{sec:method}

\subsection{Dispersion from a nonzero photon mass}
It is evidently clear from Equation~(\ref{eq2}) that two massive photons with different
frequencies ($\nu_{l}<\nu_{h}$), which are emitted simultaneously from a same source at a redshift $z$,
would reach us at different times. The arrival time delay induced by the nonzero photon
mass effect can be expressed as
\begin{equation}\label{eq:tmr}
  \Delta{t_{m_{\gamma}}}=\frac{1}{2H_0}\left(\frac{m_\gamma c^2}{h}\right)^{2}\left(\nu_l^{-2}-\nu_h^{-2}\right)H_{\gamma}(z)\;,
\end{equation}
where $H_{\gamma}(z)$ is a dimensionless redshift function,
\begin{equation}\label{eq:Hr}
  H_{\gamma}(z)=\int_{0}^{z}\frac{(1+z')^{-2}dz'}{\sqrt{\Omega_{\rm m}(1+z')^{3}+\Omega_{\Lambda}}}\;.
\end{equation}

\subsection{Dispersion from the plasma effect}
Because of the dispersive nature of plasma, lower frequency radio photons would travel across
the ionized median slower than higher frequency ones \citep{2017AdSpR..59..736B}.
The arrival time difference between two photons with different frequencies ($\nu_{l}<\nu_{h}$),
which caused by the plasma effect, is described as
\begin{equation}\label{eq:tDM}
\begin{aligned}
  \Delta{t_{\rm DM}}&=\int \frac{{\rm d}l}{c} \frac{\nu^{2}_{p}}{2}\left(\nu_l^{-2}-\nu_h^{-2}\right)\\
  &=\frac{e^{2}}{8\pi^{2} m_{e}\epsilon_{0}c}\left(\nu_l^{-2}-\nu_h^{-2}\right){\rm DM_{astro}}\;,
\end{aligned}
\end{equation}
where $\nu_{p}=(n_{e}e^{2}/4\pi^{2} m_{e}\epsilon_{0})^{1/2}$ is the plasma frequency,
$n_{e}$ is the electron number density, $m_{e}$ and $e$ are the mass and charge of an electron,
respectively, and $\epsilon_{0}$ is the permittivity. Here ${\rm DM_{astro}}$ is defined as
the integrated electron number density along the line of sight, i.e.,
${\rm DM_{astro}}\equiv\int n_{e}{\rm d}l$. In a cosmological setting,
${\rm DM_{astro}}\equiv\int n_{e,z}(1+z)^{-1}{\rm d}l$, where $n_{e,z}$ is the rest-frame
number density of electrons and $z$ is the redshift \citep{2014ApJ...783L..35D}.

For a cosmic source, the $\rm DM_{astro}$ should include the following components:
\begin{equation}\label{eq:DMastro}
  \rm DM_{astro}=DM_{MW}+DM_{MWhalo}+DM_{IGM}+\frac{DM_{host}}{1+\emph{z}}\;,
\end{equation}
where $\rm DM_{MW}$, $\rm DM_{MWhalo}$, $\rm DM_{IGM}$, and $\rm DM_{host}$ represent the
DM contributed by the Milky Way ionized interstellar medium, the Milky Way
hot gaseous halo, the intergalactic medium (IGM), and the host galaxy (including
the host galaxy interstellar medium and source environment), respectively.
The cosmological redshift factor, $1+z$, converts the rest-frame $\rm DM_{host}$
to the measured value in the observer frame \citep{2014ApJ...783L..35D}.

\subsection{Methodology}
\label{subsec:likelihood}
As described above, the radio signals of FRBs are known to arrive with a frequency-dependent
delay in time of the $1/\nu^{2}$ behavior, which are expected from both the line-of-sight
free electron content and mass effects on photon propagation (see Equations~(\ref{eq:tmr})
and (\ref{eq:tDM})). In our analysis, we suppose that the observed time delay is attributed to
both the plasma and nonzero photon mass effects, i.e.,
\begin{equation}\label{tobs}
\Delta{t_{\rm obs}}=\Delta{t_{\rm DM}}+\Delta{t_{m_{\gamma}}}\;.
\end{equation}
In practice, the observed DM of an FRB, $\rm DM_{obs}$, is directly obtained from the fitting
of the $\nu^{-2}$ form of the observed time delay. In our analysis, it thus equals to
\begin{equation}\label{eq:DMobs}
{\rm DM_{obs}}={\rm DM_{astro}}+{\rm DM_{\gamma}}\;,
\end{equation}
where $\rm DM_{\gamma}$ represents the ``effective DM'' arises from a nonzero photon mass
\citep{2017PhRvD..95l3010S},
\begin{equation}\label{DMr}
{\rm DM_{\gamma}}\equiv\frac{4\pi^2m_e\epsilon_{0}c^5}{h^2e^2}\frac{H_{\gamma}(z)}{H_0}m_\gamma^2\;.
\end{equation}

\begin{table}
\begin{center}
\caption[]{Redshifts and Dispersion Measures of Nine Localized FRBs.}\label{Tab:1}
%%Please Capitalize the First Letter of Each Notional Word in table's caption
 \begin{tabular}{lcccc}
  \hline\noalign{\smallskip}
FRB Name &  $z$      & ${\rm DM_{obs}}$ & ${\rm DM_{MW}}$   & References  \\
         &           & (pc ${\rm cm^{-3}}$)  & (pc ${\rm cm^{-3}}$) & \\
  \hline\noalign{\smallskip}
FRB 121102	&	0.19273	&$	558.1	\pm 	3.3	$&	188	&	 (1), (2), (3)\\
FRB 180916.J0158+65	&	0.0337	&$	348.76	\pm 	0.1	$&	199	& (4)\\
FRB 180924	&	0.3214	&$	361.42	\pm 	0.06	$&	40.5	& (5)\\
FRB 181112	&	0.4755	&$	589.27	\pm 	0.03	$&	102	&	(6)\\
FRB 190102	&	0.291	&$	363.6	\pm 	0.3	$&	57.3	&	 (7)\\
FRB 190523	&	0.66	&$	760.8	\pm 	0.6	$&	37	&	 (8)\\
FRB 190608	&	0.1178	&$	338.7	\pm 	0.5	$&	37.2	&	 (7)\\
FRB 190611	&	0.378	&$	321.4	\pm 	0.2	$&	57.83	&	 (7)\\
FRB 190711	&	0.522	&$	593.1	\pm 	0.4	$&	56.4	&	 (7)\\
  \noalign{\smallskip}\hline
\end{tabular}
\tablecomments{0.85\textwidth}{The References are (1) \cite{2016Natur.531..202S}; (2) \cite{2017Natur.541...58C};
(3) \cite{2017ApJ...834L...7T}; (4) \cite{2020Natur.577..190M}; (5) \cite{2019Sci...365..565B};
(6) \cite{2019Sci...366..231P}; (7) \cite{2020Natur.581..391M}; (8) \cite{2019Natur.572..352R}.}
\end{center}
\end{table}

\begin{figure}
   \centering
   \includegraphics[width=0.6\textwidth, angle=0]{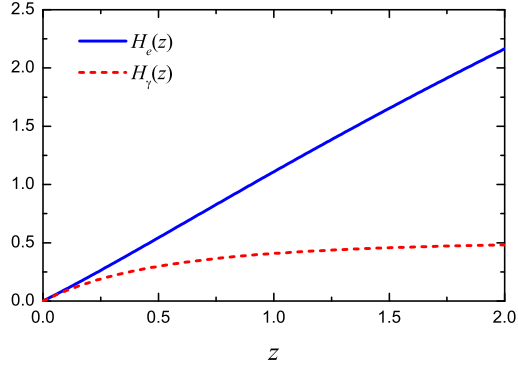}
   \vskip-0.2in
   \caption{Dependence of the two dimensionless functions, $H_{e}(z)$ and $H_{\gamma}(z)$, on the redshift $z$. }
   \label{Fig1}
   \end{figure}

The measurements of $\rm DM_{obs}$ and their corresponding uncertainties $\sigma_{\rm obs}$
for all nine localized FRBs are presented in Table~\ref{Tab:1}. These $\rm DM_{obs}$
values were determined by maximizing the peak signal-to-noise ratios.\footnote{
The frequency-dependent structure of subpulses in FRB 121102 complicates the determination of
$\rm DM_{obs}$, \cite{2019ApJ...876L..23H} argued that it is appropriate to use a $\rm DM_{obs}$
metric that maximizes structure in the frequency-averaged pulse profile, as opposed to peak
signal-to-noise, and found $\rm DM_{obs}=560.57\pm0.07$ pc ${\rm cm^{-3}}$ at MJD 57,644.
To test the dependence of our analysis on the $\rm DM_{obs}$ determination metric, we also
carried out a parallel comparative analysis using the $\rm DM_{obs}$ determination from the structure-maximizing
metric \citep{2019ApJ...876L..23H}, and found that the adoption of a different $\rm DM_{obs}$
determination metric has a very tiny influence on the photon mass limit.}
In order to identify $\rm DM_{\gamma}$ as radical as a massive photon effect from $\rm DM_{obs}$,
one needs to know the different DM contributions in Equation~(\ref{eq:DMastro}).
For a well-localized FRB, the DM due to the Milky Way, $\rm DM_{MW}$, can be well estimated
from the Galactic electron density models. For FRB 181112, its $\rm DM_{MW}$ value is derived
based on the Galactic electron density model of \cite{2019Sci...366..231P}. For the other
eight FRBs, we adopt the NE2001 model for $\rm DM_{MW}$ \citep{2002astro.ph..7156C}.
Table~\ref{Tab:1} also contains the estimated $\rm DM_{MW}$ of nine localized FRBs, which are
available in the FRB catalog \citep{2016PASA...33...45P}. The $\rm DM_{MWhalo}$ term is not
well determined, but is estimated to be in the range of approximately 50--80 pc ${\rm cm^{-3}}$
\citep{2019MNRAS.485..648P}. Hereafter we assume $\rm DM_{MWhalo}=50$ pc ${\rm cm^{-3}}$.
However, $\rm DM_{host}$ is poorly known, which depends on the type of FRB host galaxy,
the relative orientations of the host and source, and the near-source plasma \citep{2015RAA....15.1629X}.
For a given galaxy type, the scale of $\rm DM_{host}$ along a certain line of sight is related
to the size of the galaxy and electron density, and the average electron density is proportional
to the root of the galaxy H$\alpha$ luminosity \citep{2018MNRAS.481.2320L}. As the H$\alpha$
luminosity scales with the star formation rate (SFR; \citealt{1994ApJ...435...22K,1998ApJ...498..106M}),
we model $\rm DM_{host}$ as a function of star formation history \citep{2018MNRAS.481.2320L},
$\rm DM_{host}(\emph{z})=DM_{host,0}\sqrt{SFR(\emph{z})/SFR(0)}$, where $\rm DM_{host,0}$
is the present value of $\rm DM_{host}(\emph{z}=0)$ and
${\rm SFR}(z)=\frac{0.0157+0.118z}{1+(z/3.23)^{4.66}}$ $\rm M_{\odot}$ $\rm yr^{-1}$ ${\rm Mpc^{-3}}$
is the adopted star formation history \citep{2006ApJ...651..142H,2008MNRAS.388.1487L}.
In our following likelihood estimation, $\rm DM_{host,0}$ is treated as a free parameter.
The IGM portion of DM is related to the redshift of
the source and the fraction of ionized electrons in helium and hydrogen on the propagation path.
If both helium and hydrogen are fully ionized (valid below $z\sim3$), it can be written as
\citep{2014ApJ...783L..35D}
\begin{equation}\label{eq:DMIGM}
{\rm DM_{IGM}}=\frac{21cH_{0}\Omega_{b}f_{\rm IGM}}{64\pi G m_p}H_{e}(z)\;,
\end{equation}
where $m_p$ is the mass of a proton, $f_{\rm IGM}\simeq0.83$ is the fraction of baryons in the IGM
\citep{1998ApJ...503..518F}, and $H_{e}(z)$ is the redshift-dependent dimensionless function,
\begin{equation}\label{eq:He}
  H_{e}(z)=\int_{0}^{z}\frac{(1+z')dz'}{\sqrt{\Omega_{\rm m}(1+z')^{3}+\Omega_{\Lambda}}}\;.
\end{equation}
In Figure~\ref{Fig1}, we illustrate the dependence of functions $H_{e}(z)$ and $H_{\gamma}(z)$
on the redshift $z$. It is obvious that the DM contributions from the IGM and a possible
photon mass have different redshift dependences. \cite{2016PhLB..757..548B,2017PhLB..768..326B}
pointed out that the different behavior of the two redshift functions can be used to break
parameter degeneracy and to improve the sensitivity to the photon mass when a few FRB redshifts
are available (see also \citealt{2017AdSpR..59..736B,2017PhRvD..95l3010S}). For now, nine FRB
redshifts have been measured, which might make this point come true.

For a set of nine localized FRBs, a combined limit on the photon mass $m_{\gamma}$ can be obtained
by maximizing the likelihood function:
\begin{equation}\label{eq:likelihood}
\begin{aligned}
\mathcal{L} = &\prod_{i}
\frac{1}{\sqrt{2\pi}\sigma_{{\rm tot},i}}\\
&\times\exp\left\{-\,\frac{\left[{\rm DM_{obs,\emph{i}}-DM_{astro,\emph{i}}-DM_{\gamma}}(m_{\gamma},z_i)\right]^{2}}
{2\sigma_{{\rm tot},i}^{2}}\right\}\;,
\end{aligned}
\end{equation}
where $\sigma_{\rm tot}^2$ is the total variance of each FRB,
\begin{equation}
\sigma_{\rm tot}^{2}=\sigma_{\rm obs}^{2}+\sigma_{\rm MW}^{2}+\sigma_{\rm MWhalo}^{2}+\sigma_{\rm int}^{2}\;,
\end{equation}
where $\sigma_{\rm obs}$, $\sigma_{\rm MW}$, and $\sigma_{\rm MWhalo}$ are the uncertainties of
$\rm DM_{obs}$, $\rm DM_{MW}$, and $\rm DM_{MWhalo}$, respectively, and $\sigma_{\rm int}$
is the global intrinsic scatter. The ATNF pulsar catalog shows that the average uncertainty of
$\rm DM_{MW}$ for sources at high Galactic latitude ($|b|>10^{\circ}$) is about 10 pc $\rm cm^{-3}$
\citep{2005AJ....129.1993M}, and we take this value as $\sigma_{\rm MW}$. Recent observations
imply that the Galactic halo contributes between 50--80 pc $\rm cm^{-3}$ \citep{2019MNRAS.485..648P}.
Here we adopt the possible range of $\rm DM_{MWhalo}$ as its uncertainty, i.e.,
$\sigma_{\rm MWhalo}=30$ pc $\rm cm^{-3}$. To be conservative, we introduce an additional
free parameter $\sigma_{\rm int}$ to account for the intrinsic scatter that might
originate from the large IGM fluctuation and the diversity of host galaxy contribution.
In this case, the free parameters are $\rm DM_{host,0}$, $\sigma_{\rm int}$, and
the photon mass $m_{\gamma}$.

\section{Combined Photon Mass Limit from Nine Localized FRBs}
\label{sec:result}
\begin{figure}
   \centering
   \includegraphics[width=0.6\textwidth, angle=0]{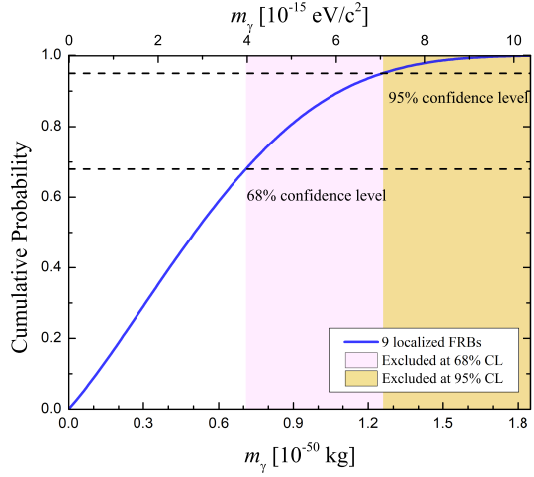}
   \vskip-0.3in
   \caption{Cumulative posterior probability distribution for the photon mass
   $m_{\gamma}$ derived from nine localized FRBs. The shadowed areas represent
   the excluded values of $m_{\gamma}$ at 68\% and 95\% confidence levels.}
   \label{Fig2}
   \end{figure}

We use the Python Markov Chain Monte Carlo module, EMCEE \citep{2013PASP..125..306F},
to obtain the posterior distributions of the three parameters ($m_{\gamma}$, $\rm DM_{host,0}$, and
$\sigma_{\rm int}$) from nine FRBs with redshift measurements. The marginalized accumulative
probability distribution of the photon mass $m_{\gamma}$ is displayed in Figure~\ref{Fig2}.
The 68\% and 95\% confidence-level upper limits on $m_{\gamma}$ are
\begin{equation}
m_{\gamma}\leq7.1\times10^{-51}\;{\rm kg}\simeq4.0\times10^{-15}\; {\rm eV}/c^{2}
\end{equation}
and
\begin{equation}
m_{\gamma}\leq1.3\times10^{-50}\; {\rm kg}\simeq7.3\times10^{-15}\; {\rm eV}/c^{2}\;,
\end{equation}
respectively. These results are comparable with or represent sensitivities improved by
a factor of 7-fold over existing photon mass limits from the single FRBs \citep{2016ApJ...822L..15W,2016PhLB..757..548B,2017PhLB..768..326B,2019ApJ...882L..13X}.

\begin{figure}
   \centering
   \includegraphics[width=0.6\textwidth, angle=0]{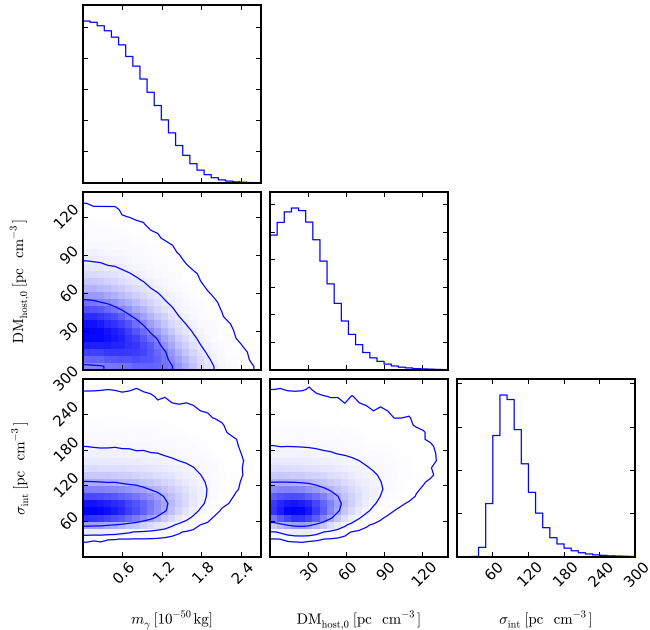}
   \vskip-0.2in
   \caption{1D and 2D marginalized probability distributions with $1-3\sigma$
    confidence contours for the photon mass $m_{\gamma}$ and the parameters
    $\rm DM_{host,0}$ and $\sigma_{\rm int}$, using the combination of nine localized FRBs.}
   \label{Fig3}
   \end{figure}

In Figure~\ref{Fig3}, we show the 1D and 2D marginalized posterior probability distributions
with $1-3\sigma$ confidence regions for the parameters. We find that the mean value of DM
contributed from the local host galaxy is estimated to be $\rm DM_{host,0}=27.0^{+23.3}_{-17.6}$
pc $\rm cm^{-3}$ at 68\% confidence level, which is well consistent with those
inferred from observations of localized FRBs \citep{2019Sci...365..565B,2019Natur.572..352R}.
Meanwhile, an estimation for the global intrinsic scatter arises from both the large IGM fluctuation
and the diversity of host galaxy contribution, $\sigma_{\rm int}=92.5^{+36.4}_{-23.7}$
pc $\rm cm^{-3}$, is obtained.

\section{Conclusion and Discussions}
\label{sec:summary}
The rest mass of the photon, $m_{\gamma}$, can be effectively constrained by measuring
the frequency-dependent time delays of radio waves from distant astrophysical sources.
Particularly, extragalactic FRBs are well suited for these studies because they are
short-duration radio transients involving long cosmological propagation distances. However,
there is a similar frequency-dependent dispersion expected from both the plasma and
possible photon mass effects on photon propagation. A key challenge in this method of
constraining $m_{\gamma}$, therefore, is to distinguish the dispersions from the plasma
and photon mass. Most previous limits were based on the dispersion measure (DM) of a single
FRB, in which the degeneracy of these two dispersions can not be well handled
\citep{2016ApJ...822L..15W,2016PhLB..757..548B,2017PhLB..768..326B}.
In view of their different redshift dependences (see Equations~(\ref{eq:Hr}) and (\ref{eq:He})),
it has been suggested that the DM contributions from the IGM and photon mass could
in principle be disentangled by more FRBs with different redshift measurements, enabling
the sensitivity to $m_{\gamma}$ to be improved
\citep{2016PhLB..757..548B,2017PhLB..768..326B,2017AdSpR..59..736B}.

Recently, nine FRBs with redshift measurements have been reported. These provide a good
opportunity not only to disentangle the degeneracy problem but also to place more stringent
limits on the photon mass. In this work, we suppose that the observed DM of an FRB, $\rm DM_{obs}$,
is attributed to both the plasma and nonzero photon mass effects. Using the measurements of
$\rm DM_{obs}$ and $z$ of the nine FRBs, we place a combined limit on the photon mass at
68\% (95\%) confidence level, i.e., $m_{\gamma}\leq7.1\times10^{-51}\;{\rm kg}$, or equivalently
$m_{\gamma}\leq4.0\times10^{-15}\; {\rm eV}/c^{2}$ ($m_{\gamma}\leq1.3\times10^{-50}\;{\rm kg}$,
or equivalently $m_{\gamma}\leq7.3\times10^{-15}\; {\rm eV}/c^{2}$), which is as good as or
represents a factor of 7 improvement over the results previously obtained from the single FRBs.
Moreover, a reasonable estimation for
the mean value of DM contributing from host galaxies is simultaneously achieved,
i.e., $\rm DM_{host,0}=27.0^{+23.3}_{-17.6}$ pc $\rm cm^{-3}$.
Previously, \cite{2017PhRvD..95l3010S} derived a combined limit of $m_{\gamma}\leq8.7\times10^{-51}\;{\rm kg}$
at 68\% confidence level from a catalog of 21 FRBs (20 of them without redshift measurement),
where uninformative prior is made to the unknown redshift. With another prior of $z$ that
traces the SFR, \cite{2017PhRvD..95l3010S} obtained a combined limit of $7.5\times10^{-51}\;{\rm kg}$
at 68\% confidence level, showing that changes in the prior of redshift can lead to slight
difference. The precision of our constraint from nine localized FRBs is comparable to these,
while avoiding potential bias from the prior of redshift.

More FRBs with redshift measurements are essential for using this method presented here
to constrain the photon rest mass. It is encouraging that the Deep Synoptic Array 10-dish
prototype (DSA-10) will gradually be replaced with the ongoing 110-dish DSA in 2021 and
the ultimate 2000-dish DSA in 2026, which are expected to detect and localize over 100
and $10^4$ FRBs per year \citep{2019BAAS...51g.255H,2019MNRAS.489..919K}. With the rapid
progress in localizing FRBs, the photon mass limits will be further improved.

%by modelling the DM contribution
%from a host galaxy, $\rm DM_{host}$, as a function of star formation history, a reasonable
%estimation for the present value of $\rm DM_{host}$ is simultaneously achieved, i.e., $\rm DM_{host,0}=43.9^{+24.1}_{-22.8}$ pc $\rm cm^{-3}$.

%%%%%%%%%%%%%%%%%%%%%%%%%%%%%%%%%%%%%%%%%%%%%%%%%%%%%%%%%%%%%%%%%%%%%%%%%%

\normalem
\begin{acknowledgements}
We are grateful to the anonymous referee for helpful comments.
This work is partially supported by the National Natural Science Foundation of China
(grant Nos. 11673068, 11725314, and U1831122), the Youth Innovation Promotion
Association (2017366), the Key Research Program of Frontier Sciences (grant Nos. QYZDB-SSW-SYS005
and ZDBS-LY-7014), and the Strategic Priority Research Program ``Multi-waveband gravitational wave universe''
(grant No. XDB23000000) of Chinese Academy of Sciences.
\end{acknowledgements}

%\bibliographystyle{raa}
%\bibliography{bibtex}

\begin{thebibliography}{62}
\providecommand\natexlab[1]{#1}
\providecommand\JournalTitle[1]{#1}

\bibitem[{Accioly} \& {Paszko}(2004)]{2004PhRvD..69j7501A}
{Accioly}, A., \& {Paszko}, R. 2004, \prd, 69, 107501

\bibitem[{Bannister} {et~al.}(2019)]{2019Sci...365..565B}
{Bannister}, K.~W., {Deller}, A.~T., {Phillips}, C., {et~al.} 2019, Science,
  365, 565

\bibitem[{Bentum} {et~al.}(2017)]{2017AdSpR..59..736B}
{Bentum}, M.~J., {Bonetti}, L., \& {Spallicci}, A. r. D.~A.~M. 2017, Advances
  in Space Research, 59, 736

\bibitem[{Bonetti} {et~al.}(2016)]{2016PhLB..757..548B}
{Bonetti}, L., {Ellis}, J., {Mavromatos}, N.~E., {et~al.} 2016, Physics Letters
  B, 757, 548

\bibitem[{Bonetti} {et~al.}(2017)]{2017PhLB..768..326B}
{Bonetti}, L., {Ellis}, J., {Mavromatos}, N.~E., {et~al.} 2017, Physics Letters
  B, 768, 326

\bibitem[{Chatterjee} {et~al.}(2017)]{2017Natur.541...58C}
{Chatterjee}, S., {Law}, C.~J., {Wharton}, R.~S., {et~al.} 2017, \nat, 541, 58

\bibitem[{Chernikov} {et~al.}(1992)]{1992PhRvL..68.3383C}
{Chernikov}, M.~A., {Gerber}, C.~J., {Ott}, H.~R., \& {Gerber}, H.~J. 1992,
  \prl, 68, 3383

\bibitem[{Chibisov}(1976)]{1976UsFiN.119..551C}
{Chibisov}, G.~V. 1976, Uspekhi Fizicheskikh Nauk, 119, 551

\bibitem[{Cordes} \& {Lazio}(2002)]{2002astro.ph..7156C}
{Cordes}, J.~M., \& {Lazio}, T.~J.~W. 2002, arXiv:astro-ph/0207156

\bibitem[{Davis} {et~al.}(1975)]{1975PhRvL..35.1402D}
{Davis}, L., J., {Goldhaber}, A.~S., \& {Nieto}, M.~M. 1975, \prl, 35, 1402

\bibitem[{de Broglie}(1922)]{deb22}
{de Broglie}, L. 1922, J.~Phys.~Radium, 3, 422

\bibitem[{Deng} \& {Zhang}(2014)]{2014ApJ...783L..35D}
{Deng}, W., \& {Zhang}, B. 2014, \apjl, 783, L35

\bibitem[{Foreman-Mackey} {et~al.}(2013)]{2013PASP..125..306F}
{Foreman-Mackey}, D., {Hogg}, D.~W., {Lang}, D., \& {Goodman}, J. 2013, \pasp,
  125, 306

\bibitem[{Fukugita} {et~al.}(1998)]{1998ApJ...503..518F}
{Fukugita}, M., {Hogan}, C.~J., \& {Peebles}, P.~J.~E. 1998, \apj, 503, 518

\bibitem[{Goldhaber} \& {Nieto}(1971)]{1971RvMP...43..277G}
{Goldhaber}, A.~S., \& {Nieto}, M.~M. 1971, Reviews of Modern Physics, 43, 277

\bibitem[{Goldhaber} \& {Nieto}(2003)]{2003PhRvL..91n9101G}
{Goldhaber}, A.~S., \& {Nieto}, M.~M. 2003, \prl, 91, 149101

\bibitem[{Goldhaber} \& {Nieto}(2010)]{2010RvMP...82..939G}
{Goldhaber}, A.~S., \& {Nieto}, M.~M. 2010, Reviews of Modern Physics, 82, 939

\bibitem[{Hallinan} {et~al.}(2019)]{2019BAAS...51g.255H}
{Hallinan}, G., {Ravi}, V., {Weinreb}, S., {et~al.} 2019, in \baas, Vol.~51,
  255

\bibitem[{Hessels} {et~al.}(2019)]{2019ApJ...876L..23H}
{Hessels}, J.~W.~T., {Spitler}, L.~G., {Seymour}, A.~D., {et~al.} 2019, \apjl,
  876, L23

\bibitem[{Hopkins} \& {Beacom}(2006)]{2006ApJ...651..142H}
{Hopkins}, A.~M., \& {Beacom}, J.~F. 2006, \apj, 651, 142

\bibitem[{Kennicutt} {et~al.}(1994)]{1994ApJ...435...22K}
{Kennicutt}, Robert~C., J., {Tamblyn}, P., \& {Congdon}, C.~E. 1994, \apj, 435,
  22

\bibitem[{Kocz} {et~al.}(2019)]{2019MNRAS.489..919K}
{Kocz}, J., {Ravi}, V., {Catha}, M., {et~al.} 2019, \mnras, 489, 919

\bibitem[{Kouwn} {et~al.}(2016)]{2016PhRvD..93h3012K}
{Kouwn}, S., {Oh}, P., \& {Park}, C.-G. 2016, \prd, 93, 083012

\bibitem[{Lakes}(1998)]{1998PhRvL..80.1826L}
{Lakes}, R. 1998, \prl, 80, 1826

\bibitem[{Li}(2008)]{2008MNRAS.388.1487L}
{Li}, L.-X. 2008, \mnras, 388, 1487

\bibitem[{Lovell} {et~al.}(1964)]{1964Natur.202..377L}
{Lovell}, B., {Whipple}, F.~L., \& {Solomon}, L.~H. 1964, \nat, 202, 377

\bibitem[{Lowenthal}(1973)]{1973PhRvD...8.2349L}
{Lowenthal}, D.~D. 1973, \prd, 8, 2349

\bibitem[{Luo} {et~al.}(2003{\natexlab{a}})]{2003PhRvL..91n9102L}
{Luo}, J., {Tu}, L.-C., {Hu}, Z.-K., \& {Luan}, E.-J. 2003{\natexlab{a}}, \prl,
  91, 149102

\bibitem[{Luo} {et~al.}(2003{\natexlab{b}})]{2003PhRvL..90h1801L}
{Luo}, J., {Tu}, L.-C., {Hu}, Z.-K., \& {Luan}, E.-J. 2003{\natexlab{b}}, \prl,
  90, 081801

\bibitem[{Luo} {et~al.}(2018)]{2018MNRAS.481.2320L}
{Luo}, R., {Lee}, K., {Lorimer}, D.~R., \& {Zhang}, B. 2018, \mnras, 481, 2320

\bibitem[{Macquart} {et~al.}(2020)]{2020Natur.581..391M}
{Macquart}, J.~P., {Prochaska}, J.~X., {McQuinn}, M., {et~al.} 2020, \nat, 581,
  391

\bibitem[{Madau} {et~al.}(1998)]{1998ApJ...498..106M}
{Madau}, P., {Pozzetti}, L., \& {Dickinson}, M. 1998, \apj, 498, 106

\bibitem[{Manchester} {et~al.}(2005)]{2005AJ....129.1993M}
{Manchester}, R.~N., {Hobbs}, G.~B., {Teoh}, A., \& {Hobbs}, M. 2005, \aj, 129,
  1993

\bibitem[{Marcote} {et~al.}(2020)]{2020Natur.577..190M}
{Marcote}, B., {Nimmo}, K., {Hessels}, J.~W.~T., {et~al.} 2020, \nat, 577, 190

\bibitem[{Okun}(2006)]{2006AcPPB..37..565O}
{Okun}, L.~B. 2006, Acta Physica Polonica B, 37, 565

\bibitem[{Pani} {et~al.}(2012)]{2012PhRvL.109m1102P}
{Pani}, P., {Cardoso}, V., {Gualtieri}, L., {Berti}, E., \& {Ishibashi}, A.
  2012, \prl, 109, 131102

\bibitem[{Petroff} {et~al.}(2016)]{2016PASA...33...45P}
{Petroff}, E., {Barr}, E.~D., {Jameson}, A., {et~al.} 2016, \pasa, 33, e045

\bibitem[{Planck Collaboration} {et~al.}(2018)]{2018arXiv180706209P}
{Planck Collaboration}, {Aghanim}, N., {Akrami}, Y., {et~al.} 2018, arXiv
  e-prints, arXiv:1807.06209

\bibitem[{Proca}(1936)]{pro36}
{Proca}, A. 1936, J.~Phys.~Radium, 7, 347

\bibitem[{Prochaska} \& {Zheng}(2019)]{2019MNRAS.485..648P}
{Prochaska}, J.~X., \& {Zheng}, Y. 2019, \mnras, 485, 648

\bibitem[{Prochaska} {et~al.}(2019)]{2019Sci...366..231P}
{Prochaska}, J.~X., {Macquart}, J.-P., {McQuinn}, M., {et~al.} 2019, Science,
  366, 231

\bibitem[{Ravi} {et~al.}(2019)]{2019Natur.572..352R}
{Ravi}, V., {Catha}, M., {D'Addario}, L., {et~al.} 2019, \nat, 572, 352

\bibitem[{Retin{\`o}} {et~al.}(2016)]{2016APh....82...49R}
{Retin{\`o}}, A., {Spallicci}, A. D.~A.~M., \& {Vaivads}, A. 2016,
  Astroparticle Physics, 82, 49

\bibitem[{Ryutov}(1997)]{1997PPCF...39...73R}
{Ryutov}, D.~D. 1997, Plasma Physics and Controlled Fusion, 39, A73

\bibitem[{Ryutov}(2007)]{2007PPCF...49..429R}
{Ryutov}, D.~D. 2007, Plasma Physics and Controlled Fusion, 49, B429

\bibitem[{Schaefer}(1999)]{1999PhRvL..82.4964S}
{Schaefer}, B.~E. 1999, \prl, 82, 4964

\bibitem[{Shao} \& {Zhang}(2017)]{2017PhRvD..95l3010S}
{Shao}, L., \& {Zhang}, B. 2017, \prd, 95, 123010

\bibitem[{Spavieri} {et~al.}(2011)]{2011EPJD...61..531S}
{Spavieri}, G., {Quintero}, J., {Gillies}, G.~T., \& {Rodr{\'\i}guez}, M. 2011,
  European Physical Journal D, 61, 531

\bibitem[{Spitler} {et~al.}(2016)]{2016Natur.531..202S}
{Spitler}, L.~G., {Scholz}, P., {Hessels}, J.~W.~T., {et~al.} 2016, \nat, 531,
  202

\bibitem[{Tendulkar} {et~al.}(2017)]{2017ApJ...834L...7T}
{Tendulkar}, S.~P., {Bassa}, C.~G., {Cordes}, J.~M., {et~al.} 2017, \apjl, 834,
  L7

\bibitem[{Tu} {et~al.}(2005)]{2005RPPh...68...77T}
{Tu}, L.-C., {Luo}, J., \& {Gillies}, G.~T. 2005, Reports on Progress in
  Physics, 68, 77

\bibitem[{Vedantham} {et~al.}(2016)]{2016ApJ...824L...9V}
{Vedantham}, H.~K., {Ravi}, V., {Mooley}, K., {et~al.} 2016, \apjl, 824, L9

\bibitem[{Warner} \& {Nather}(1969)]{1969Natur.222..157W}
{Warner}, B., \& {Nather}, R.~E. 1969, \nat, 222, 157

\bibitem[{Wei} \& {Wu}(2018)]{2018JCAP...07..045W}
{Wei}, J.-J., \& {Wu}, X.-F. 2018, \jcap, 2018, 045

\bibitem[{Wei} {et~al.}(2017)]{2017RAA....17...13W}
{Wei}, J.-J., {Zhang}, E.-K., {Zhang}, S.-B., \& {Wu}, X.-F. 2017, Research in
  Astronomy and Astrophysics, 17, 13

\bibitem[{Williams} {et~al.}(1971)]{1971PhRvL..26..721W}
{Williams}, E.~R., {Faller}, J.~E., \& {Hill}, H.~A. 1971, \prl, 26, 721

\bibitem[{Williams} \& {Berger}(2016)]{2016ApJ...821L..22W}
{Williams}, P.~K.~G., \& {Berger}, E. 2016, \apjl, 821, L22

\bibitem[{Wu} {et~al.}(2016)]{2016ApJ...822L..15W}
{Wu}, X.-F., {Zhang}, S.-B., {Gao}, H., {et~al.} 2016, \apjl, 822, L15

\bibitem[{Xing} {et~al.}(2019)]{2019ApJ...882L..13X}
{Xing}, N., {Gao}, H., {Wei}, J.-J., {et~al.} 2019, \apjl, 882, L13

\bibitem[{Xu} \& {Han}(2015)]{2015RAA....15.1629X}
{Xu}, J., \& {Han}, J.~L. 2015, Research in Astronomy and Astrophysics, 15,
  1629

\bibitem[{Yang} \& {Zhang}(2017)]{2017ApJ...842...23Y}
{Yang}, Y.-P., \& {Zhang}, B. 2017, \apj, 842, 23

\bibitem[{Zhang} {et~al.}(2016)]{2016JHEAp..11...20Z}
{Zhang}, B., {Chai}, Y.-T., {Zou}, Y.-C., \& {Wu}, X.-F. 2016, Journal of High
  Energy Astrophysics, 11, 20

\end{thebibliography}

\end{document}